\documentclass[aps,prl,twocolumn,showpacs,floatfix,superscriptaddress]{revtex4-1}
\usepackage{amssymb,amsmath,bm}
\usepackage{graphicx}
\usepackage[colorlinks]{hyperref}

\begin{document}

\title{Charged-current weak interaction processes in hot and dense
  matter and its impact on the spectra of neutrinos emitted from
  proto-neutron star cooling}

\author{G.~Mart{\'i}nez-Pinedo}
\affiliation{Institut f{\"u}r Kernphysik, Technische Universit{\"a}t
  Darmstadt, Schlossgartenstra{\ss}e 2, 64289 Darmstadt, Germany} 
\affiliation{GSI Helmholtzzentrum f\"ur Schwerioneneforschung,
  Planckstra{\ss}e~1, 64291 Darmstadt, Germany} 

\author{T.~Fischer}
\affiliation{GSI Helmholtzzentrum f\"ur Schwerioneneforschung,
  Planckstra{\ss}e~1, 64291 Darmstadt, Germany} 
\affiliation{Institut f{\"u}r Kernphysik, Technische Universit{\"a}t
  Darmstadt, Schlossgartenstra{\ss}e 2, 64289 Darmstadt, Germany} 

\author{A. Lohs}
\affiliation{Institut f{\"u}r Kernphysik, Technische Universit{\"a}t
  Darmstadt, Schlossgartenstra{\ss}e 2, 64289 Darmstadt, Germany} 

\author{L. Huther}
\affiliation{Institut f{\"u}r Kernphysik, Technische Universit{\"a}t
  Darmstadt, Schlossgartenstra{\ss}e 2, 64289 Darmstadt, Germany}

\begin{abstract}
  We have performed three-flavor Boltzmann neutrino transport
  radiation hydrodynamics simulations covering a period of 3~s after
  the formation of a protoneutron star in a core-collapse supernova
  explosion. Our results show that a treatment of charged-current
  neutrino interactions in hot and dense matter as suggested
  by~\citeauthor{Reddy.Prakash.Lattimer:1998}
  [Phys. Rev. D~\textbf{58}, 013009 (1998)] has a strong impact on the
  luminosities and spectra of the emitted neutrinos. When compared
  with simulations that neglect mean field effects on the neutrino
  opacities, we find that the luminosities of all neutrino flavors are
  reduced while the spectral differences between electron neutrino and
  antineutrino are increased.  Their magnitude depends on the equation
  of state and in particular on the symmetry energy at sub-nuclear
  densities. These modifications reduce the proton-to-nucleon ratio of
  the outflow, increasing slightly their entropy.  They are expected
  to have a substantial impact on the nucleosynthesis in
  neutrino-driven winds, even though they do not result in conditions
  that favor an r-process.  Contrarily to previous findings, our
  simulations show that the spectra of electron neutrinos remain
  substantially different from those of other (anti)neutrino flavors
  during the entire deleptonization phase of the protoneutron star.
  The obtained luminosity and spectral changes are also expected to
  have important consequences for neutrino flavor oscillations and
  neutrino detection on Earth.
\end{abstract}

\date{\today}


\maketitle

Supernova explosions of massive stars are related to the birth of
neutron stars due to the collapse of the stellar core at the end of
stellar evolution~\cite{Janka.Langanke.ea:2007}.  The detection of
neutrinos from SN1987A~\cite{Hirata.Kajita.ea:1987,*Koshiba:1992}
confirmed that the $\approx 3\times10^{53}$~ergs of gravitational
energy gained by the core collapse are emitted as neutrino radiation
on time scales of tens of seconds, during which the central
protoneutron star (PNS) cools, deleptonizes and contracts.  In the
delayed neutrino-heating explosion
mechanism~\cite{Bethe.Wilson:1985,Janka.Langanke.ea:2007}, neutrinos
also transport energy from the nascent PNS to the stalled bounce
shock. This mechanism remains the most viable scenario to explain
supernova explosions as confirmed by recent two-dimensional
simulations~\cite{Mueller.Janka.Marek:2012}.  Once the explosion sets
in, the continuous emission of neutrinos from the PNS drives a
low-mass outflow known as neutrino-driven
wind~\cite{duncan.shapiro.wasserman:1986} that is currently considered
the favored site for the productions of elements heavier than iron
(e.g.~\cite{Qian:2003}).  As neutrinos travel through the stellar
mantle, they can suffer flavor
oscillations~\cite{Duan.Fuller.Qian:2010}, contribute to the
nucleosynthesis of several rare
isotopes~\cite{Woosley.Hartmann.ea:1990,*Heger.Kolbe.ea:2005} and even
drive an $r$ process in the He-shell of metal-poor
stars~\cite{Banerjee.Haxton.Qian:2011} before they are finally
detected on Earth.

Accounting for all aspects discussed above requires the knowledge of
the spectra of the neutrinos emitted during the cooling phase of the
PNS. Due to their low energies $\nu_{\mu,\tau},\bar\nu_{\mu,\tau}$
interact only via neutral-current reactions. Hence, together with the
neutron-richness of the PNS surface one expects the following
neutrino-energy hierarchy: $\langle E_{\nu_{\mu,\tau}} \rangle>\langle
E_{\bar{\nu}_e} \rangle>\langle E_{\nu_e}
\rangle$~\cite{keil.raffelt.janka:2003,Fischer.Martinez-Pinedo.ea:2012},
with $\langle E \rangle$ the average energy of the neutrino spectrum.
Early supernova
models~\cite{Woosley.Wilson.ea:1994,*Takahashi.Witti.Janka:1994}
predicted large energy differences between $\bar{\nu}_e$ and $\nu_e$
that resulted in neutron-rich ejecta as required by $r$-process
nucleosynthesis~\cite{hoffman.woosley.qian:1997}.  However, as the
treatment of neutrino transport and weak interaction processes
improved, the computed energy difference between $\bar{\nu}_e$ and
$\nu_e$ decreased and the early wind ejecta became proton
rich~\cite{Liebendoerfer.Mezzacappa.ea:2001a,*Buras.Rampp.ea:2006}. More
recently, it has been possible to perform supernova simulations based
on three-flavor Boltzmann neutrino transport for time scales of
several tens of
seconds~\cite{Huedepohl.Mueller.ea:2010,Fischer.Whitehouse.ea:2010},
covering the whole deleptonization of the PNS. These simulations
predict a continuous decrease of the energy difference between
neutrinos and antineutrinos of all flavors that became practically
indistinguishable after $\approx 10$~s. The exact value depends on
the progenitor.  The proton-richness of the ejecta increases
continuously with time and leaves the $\nu p$
process~\cite{Froehlich.Martinez-Pinedo.ea:2006,*Pruet.Hoffman.ea:2006,*Wanajo:2006}
as the only mechanism for producing elements heavier than iron in
neutrino-driven winds.

The simulations of ref.~\cite{Fischer.Whitehouse.ea:2010} have been
recently analyzed, showing that the convergence of neutrino and
antineutrino spectra at late times is due to the suppression of
charged-current processes at high
densities~\cite{Fischer.Martinez-Pinedo.ea:2012}.  This analysis was
based on a set of neutrino opacities that assume a non-interacting gas
of nucleons and nuclei.  This approximation may be valid during the
accretion phase prior to the onset of the supernova explosion when the
region from where neutrinos decouple, the neutrinospheres, is located
at relatively low densities, $\sim 10^{11}$~g~cm$^{-3}$.  However, as
the PNS cools the neutriospheres move to increasingly higher densities
where the non-interacting gas approximation breaks down. The nuclear
interaction is treated at the mean-field level in equations of state
(EoS) commonly used in core-collapse supernova
simulations~\cite{Lattimer.Swesty:1991,Shen.Toki.ea:1998a}. However,
its influence on weak interaction processes is often neglected.  In
this Letter, we show that a treatment of neutrino-matter interactions
that is consistent with the underlying EoS has a strong impact on the
spectra and luminosities of the emitted neutrinos. We discuss the
relevance for nucleosynthesis, neutrino oscillation studies and
neutrino detection.

Our study is based on the work of
Ref.~\cite{Reddy.Prakash.Lattimer:1998} where corrections to the
opacities due to strong interactions are considered at the mean-field
level.  Effects of many-body
correlations~\cite{Reddy.Prakash.ea:1999,*Burrows.Sawyer:1998,*Burrows.Sawyer:1999}
will be considered in a forthcoming publication.  They are expected to
affect the neutrino spectra at later times~\cite{Roberts.Shen.ea:2012}
than considered in the present study.  We focus on charged-current
(anti)neutrino absorption processes on neutrons and protons and the
inverse reactions: $e^- + p \rightleftarrows n + \nu_e$ and $e^+ + n
\rightleftarrows p + \bar{\nu}_e$, which are those mainly affected by
mean-field corrections.

EoS commonly used in core-collapse supernova simulations, see e.g.
Refs.~\cite{Lattimer.Swesty:1991,Shen.Toki.ea:1998a}, treat neutrons
and protons as a gas of quasi-particles that move in a
mean-field single-particle potential $U$.  Assuming non-relativistic
nucleons, which is a good approximation for densities $\rho \leq 5
\rho_0$ where $\rho_0 \approx 2.5\times 10^{14}$~g~cm$^{-3}$, the
energy momentum relation closely resembles the non-interacting
case~\cite{Reddy.Prakash.Lattimer:1998}:
\begin{equation}
\label{eq:1}
E_i(\bm{p}_i) = \frac{\bm{p}^2_i}{2m^*_i} + m_i + U_i, \quad i = n,p,
\end{equation}
with particle rest-masses $m_i$.
Both the single-particle mean-field potentials and the (Landau) effective
masses, $m^*_i$ depend on density, temperature and proton-to-nucleon ratio,
$Y_e$.
Importantly, due to the extreme neutron-rich conditions the mean-field
potentials for neutron and protons can be very different with their
relative difference $U_n - U_p$ directly related to the nuclear symmetry
energy~\cite{Reddy.Prakash.Lattimer:1998}.

In order to quantify the impact of the mean field potentials, let us
consider (anti)neutrino absorption on neutrons (protons). Assuming
zero momentum transfer, i.e.  $\bm{p}_n \approx \bm{p}_p$ (elastic
approximation), the electron(positron) and (anti)neutrino energies are
related by:
\begin{subequations}
  \label{eq:2}
  \begin{eqnarray}
    \label{eq:3}
    E_{\nu_e} &=& E_{e^-} - (m_n - m_p) - (U_n - U_p), \\
    \label{eq:4}
    E_{\bar{\nu}_e} &=& E_{e^+} +  (m_n - m_p) + (U_n - U_p).  
  \end{eqnarray}
\end{subequations}
Eqs.~(\ref{eq:3}) and~(\ref{eq:4}) show that the contribution of the
mean field potential reduces (increases) the energy of the emitted
neutrinos (antineutrinos).  The energy difference between neutrinos
and antineutrinos is increased by an amount $2(U_n - U_p)$.  The
opacity, or inverse mean-free path, for (anti)neutrino absorption also
changes.  In the elastic approximation and assuming extreme
relativistic electrons, the opacity for neutrino absorption
becomes~\cite{Bruenn:1985,Reddy.Prakash.Lattimer:1998}:
\begin{eqnarray}
\label{eq:5}
\chi(E_{\nu_e}) &=& 
\frac{G_F^2 V_{ud}^2}{\pi (\hbar c)^4} (g_V^2+3 g_A^2) \cdot
\nonumber \\
&& E_{e}^2 [1-f_e(E_{e})] \frac{n_n - n_p}{1-e^{\beta(\eta_p -U_p - \eta_n+U_n)}},
\end{eqnarray}
with $E_{\nu_e}$ and $E_e$ related by equation~(\ref{eq:3}).  The
emissivity, $j(E_{\nu_e})$, can be obtained from the detailed balance
condition $j(E_{\nu_e}) = \exp(-\beta(E_{\nu_e} -
\mu^{\text{eq}}_\nu)) \chi(E_{\nu_e})$, with $\mu^{\text{eq}}_\nu =
\mu_e - (\mu_n - \mu_p)$ the equilibrium neutrino chemical potential,
$\mu$ the chemical potential including rest mass and $\beta$ the
inverse temperature.  The opacity and emissivity for antineutrino
absorption are obtained exchanging neutron and proton and using
equation~(\ref{eq:4}) to relate the positron and antineutrino
energies.  In equation~(\ref{eq:5}), $G_F$ is the Fermi coupling
constant, $V_{ud}$ is the up-down entry of the
Cabibbo-Kobayashi-Maskawa matrix, $g_V$ and $g_A$ are vector and
axial-vector coupling constants and $n_{p,n}$ the number density of
protons or neutrons.  $f$ is the Fermi-Dirac distribution function and
$\eta$ is the chemical potential (without rest mass). The quantity
$\varphi = \eta - U$ represents the chemical potential for a
non-interacting gas of nucleons, that is related to the nucleon number
density by the relation:
\begin{equation}
\label{eq:6}
n = 2\int \frac{d^3\bm{p}}{(2\pi\hbar c)^3}
\frac{1}{e^{\beta\left(\frac{p^2}{2m} -\varphi\right)}+1}.  
\end{equation}
Eq.~(\ref{eq:6}) provides a method of determining the mean-field
potential, $U$, when using an EoS that does not 
provide this quantity,
e.g., the EoS of ref.~\cite{Shen.Toki.ea:1998a}.  

\begin{figure}[htb]
\centering
\includegraphics[width=\linewidth]{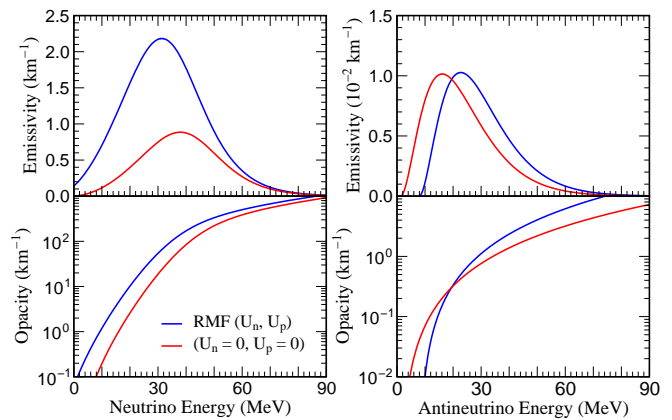}
\caption{(Color online) Opacity and emissivity for neutrino (left panels)
and antineutrino (right panels), evaluated at conditions
$\rho = 2.1\times10^{13}$~g~cm$^{-3}$, $T= 7.4$~MeV and
$Y_e=0.035$.}
\label{fig:opacity}
\end{figure}

Figure~\ref{fig:opacity} shows neutrino and antineutrino opacities and
emissivities evaluated at conditions found at the antineutrinosphere
for the 18~M$_\odot$ model of
ref.~\cite{Fischer.Martinez-Pinedo.ea:2012} at 1~s after bounce.  The
curves labeled RMF~($U_n,U_p$) include the contribution of the
mean-field potentials $U_n = -7.6$~MeV and $U_p =
-14.2$~MeV~\cite{Shen.Toki.ea:1998a}, while the curves labeled
($U_n=0,U_p=0$) assume a non-interacting gas of nucleons, i.e. neglect
the contribution of the potentials but still use chemical potentials
as given by the EoS.  Due to the presence of the mean-field potentials
the effective $Q$-value for electron capture increases with respect to
the free case producing neutrinos with substantially lower energy.
For the inverse process, neutrino absorption, the opacity is enhanced
due to the fact that the produced electron gains an energy $U_n - U_p$
reducing the final-state Pauli blocking of the electron. The situation
is completely analogous to (anti)neutrino emission and absorption on
heavy neutron-rich nuclei~\cite{Langanke.Martinez-Pinedo:2003}. Using
Eq.~(\ref{eq:5}), it can be shown that the opacity for the
non-interacting gas, $\chi_{\text{ni}}$, is related to the mean-field
opacity, $\chi_{\text{mf}}$ by $\chi_{\text{mf}}(E) = \chi_{\text{ni}}
(E+U_n-U_p)$.  This relationship produces a large enhancement of the
neutrino opacity at high densities, $\rho \approx 10^{14}$, where $U_n
- U_p \approx 50$~MeV when compared with the non-interacting
approximation used in ref.~\cite{Fischer.Martinez-Pinedo.ea:2012}. For
antineutrino absorption, due to the fact that positrons follow
Boltzmann statistics, the non-interacting emissivity and mean-field
emissivities are related by
$j_{\text{mf}}(E)=j_{\text{ni}}(E-U_n+U_p)$.  The mean-field
antineutrino opacity is larger at high densities as final-state Pauli
blocking of the neutrons becomes less efficient.

In the following, we explore the impact that a description of
opacities consistent with the EoS has on the spectra and luminosities
of the emitted neutrinos.  We have performed core-collapse supernova
simulations based on spherically symmetric radiation hydrodynamics
with three-flavor Boltzmann neutrino transport. Since our goal is to
explore the differences in neutrino energies and luminosities due to
the inclusion of mean-field potentials, we have used a low resolution
transport scheme with 12 energy bins and allowed only for radially in
and outgoing neutrinos. Despite of its limited resolution, it
reproduces the absolute values of luminosities and average energies
predicted by higher resolution
simulations~\cite{Huedepohl.Mueller.ea:2010,Fischer.Whitehouse.ea:2010}.
Table~1 of~\cite{Fischer.Martinez-Pinedo.ea:2012} list the weak
processes considered in our simulations. We use the baryonic
high-density EoS from Shen~\emph{et~al.}~\cite{Shen.Toki.ea:1998a} for
matter in nuclear statistical equilibrium (NSE) at temperatures above
0.45~MeV.  As the tabulation of Shen~\emph{et al.} does not provide
the mean-field potentials, we have computed them using
eq.~(\ref{eq:6}). In the non-NSE regime, we use the EoS of
ref.~\cite{Timmes.Arnett:1999}, which we also use for electrons,
positrons and photons in the NSE regime.  The simulations are based on
the 15~M$_\odot$ progenitor of ref.~\cite{Woosley.Heger.Weaver:2002}.
Because spherically symmetric simulations do not result in explosions
for such a massive iron-core progenitor, we enhance the neutrino
heating rates in the gain region following the scheme of
ref.~\cite{Fischer.Whitehouse.ea:2010}.  It results in the onset of
explosion at about 350~ms post bounce.  The simulations are evolved
from core collapse, through the explosion up to more than 3~seconds
after bounce.  During core collapse and post-bounce accretion phases,
the mean-field potentials, $U_n,U_p$, are only on the order of several
100~keV in the region of neutrino decoupling, which is located at
intermediate densities on the order of $10^{11}$~g~cm$^{-3}$. Hence,
the inclusion of mean-field potentials does not affect the supernova
dynamics prior to the explosion.

\begin{figure}[htp]
  \centering
  \includegraphics[width=\linewidth]{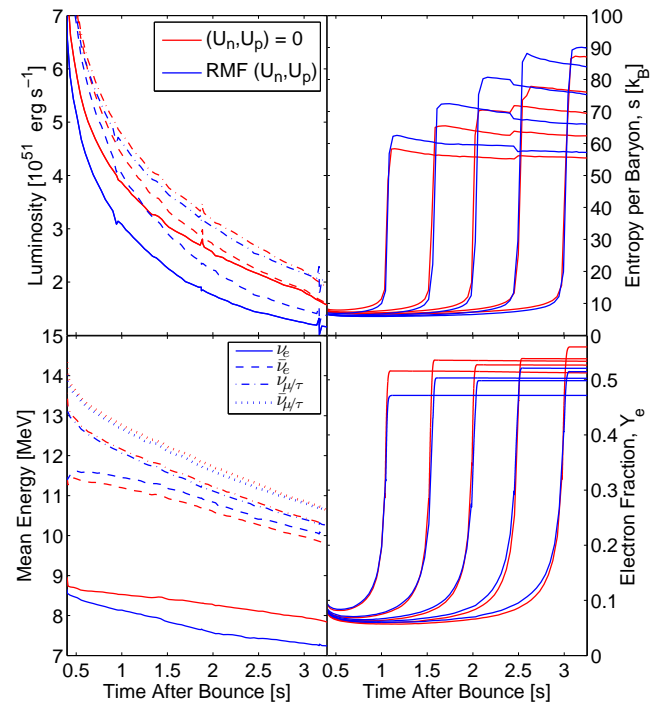}%
  \caption{(Color online) The left panel shows the neutrino luminosity
    (upper) and average energy (lower) evolution.  The right panel
    shows the evolution of the proton-to-nucleon ratio, $Y_e$, and the
    entropy per nucleon for several mass elements ejected from the PNS
    surface. The curves shown in blue use neutrino opacities computed using the
    mean-field potentials of the EoS~\cite{Shen.Toki.ea:1998a} while
    they are neglected on the red curves. \label{fig:lumin.tracer}}
\end{figure}

After the onset of the explosion  the neutrinospheres move
to increasingly higher densities reaching values of the order of
$10^{13}$~g~cm$^{-3}$.  The left panels of Fig.~\ref{fig:lumin.tracer}
show the evolution of the luminosity and average neutrino energy for
all neutrino flavors.  These observables are sampled in a co-moving
reference frame at a distance of 1000~km. Using charged-current
neutrino opacities that include the mean-field potentials slightly
reduces the luminosities for all neutrino flavors.  Moreover, as
expected from the discussion above, it enhances the differences in
luminosities and average energies between neutrinos and antineutrinos.

\begin{figure}[htb]
  \centering
  \includegraphics[width=\linewidth]{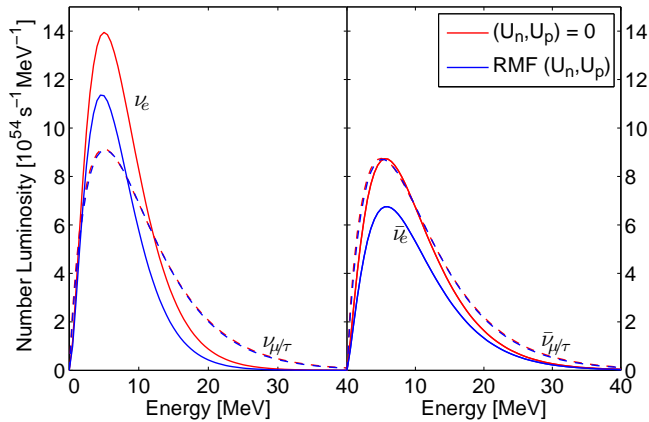}%
  \caption{(Color online) Neutrino spectra for all flavors (solid
    lines left panel: $\nu_e$, dashed lines left panel:
    $\nu_{\mu/\tau}$, solid lines right panel: $\bar\nu_e$, dashed
    lines left panel: $\bar\nu_{\mu/\tau}$), including the mean-field
    potentials (blue) and without (red).
\label{fig:spectra}}
\end{figure}

Fig.~\ref{fig:spectra} shows the different neutrino spectra for all
flavors at 3~s after bounce at a distance of 30~km outside the
neutrinospheres. At this distance neutrinos can be considered free
streaming but they have not yet been subject to collective neutrino
flavor oscillations~\cite{Duan.Fuller.Qian:2010}. These may result
in spectral swaps~\cite{Dasgupta.Dighe.ea:2009} that occur in regions
near to spectral crossings.
We expect substantially different oscillation patterns for the
spectra obtained with opacities consistent with the EoS.

The changes in electron (anti)neutrino spectra and luminosities have
important consequences for nucleosynthesis in neutrino-driven winds.
The increased difference between average energies of $\nu_e$ and
$\bar{\nu}_e$ spectra impacts the $Y_e$ of the ejected matter
resulting in slightly neutron-rich conditions for the early ejecta
(see lower-right panel of Fig.~\ref{fig:lumin.tracer}) while at later
times the ejecta become proton rich. In the simulation that neglects
the contributions of the mean-field potentials the ejecta are always
proton-rich. The decrease in $Y_e$ is not large enough to favor an
r-process but may help in the production of isotopes like $^{92}$Mo
that are only made under slightly neutron-rich
conditions~\cite{Hoffman.Woosley.ea:1996,Wanajo:2006}. Also relevant
for nucleosynthesis is the slight increase in entropy per nucleon of
the ejected material (see upper-right panel
Fig.~\ref{fig:lumin.tracer}) that can be related to the reduced
neutrino luminosities~\cite{Qian.Woosley:1996}.

We have shown that a treatment of the charged-current (anti)neutrino
opacities, that is consistent with the EoS as suggested
by~\cite{Reddy.Prakash.Lattimer:1998}, has important consequences for
the neutrino-spectra evolution during the PNS cooling phase. The most
relevant finding is an increased difference between average energies
of $\nu_e$ and $\bar{\nu}_e$ that persist during the whole simulation
time of 3~seconds after the onset of the explosion. The changes on
neutrino spectra are expected to have important consequences for
nucleosynthesis, flavor oscillations and neutrino detection on
Earth. Our results imply that not only the evolution of the neutrino
luminosities~\cite{Roberts.Shen.ea:2012} but also the spectral
differences between $\nu_e$ and $\bar{\nu}_e$ are sensitive to the
symmetry energy of nuclear matter.  Our simulations are based on
neutrino opacities computed using the elastic approximation that
neglects momentum exchange between nucleons.  They need to be extended
to consider the full kinematics~\cite{Reddy.Prakash.Lattimer:1998} of
the reaction and many-body
correlations~\cite{Reddy.Prakash.ea:1999,*Burrows.Sawyer:1998,*Burrows.Sawyer:1999}
that are expected to become important at later times than those
considered in the present study. Furthermore, it is important to
explore the sensitivity of the results to different EoS and in
particular to EoS that are consistent with recent constrains on the
nuclear symmetry energy~\cite{Lattimer.Lim:2012}.

G.M.P. is partly supported by the Deutsche Forschungsgemeinschaft
through contract SFB 634, the Helmholtz International Center for FAIR
within the framework of the LOEWE program launched by the state of
Hesse and the Helmholtz Association through the Nuclear Astrophysics
Virtual Institute (VH-VI-417).  T.F. is supported by the Swiss
National Science Foundation under project~no.~PBBSP2-133378. A.L. is
supported by the Helmholtz International Center for FAIR and GSI
Helmholtzzentrum f\"ur Schwerioneneforschung. L.H. is supported by
the Deutsche Forschungsgemeinschaft through contract SFB 634. We thank
fruitful discussions with B.~Friman, M.~Hempel, H.-Th.~Janka,
K.~Langanke, J.~M.~Lattimer, M.~Liebend\"orfer, B.~M\"uller,
F.-K.~Thielemann, and S. Typel.


\begin{thebibliography}{38}%
\makeatletter
\providecommand \@ifxundefined [1]{%
 \@ifx{#1\undefined}
}%
\providecommand \@ifnum [1]{%
 \ifnum #1\expandafter \@firstoftwo
 \else \expandafter \@secondoftwo
 \fi
}%
\providecommand \@ifx [1]{%
 \ifx #1\expandafter \@firstoftwo
 \else \expandafter \@secondoftwo
 \fi
}%
\providecommand \natexlab [1]{#1}%
\providecommand \enquote  [1]{``#1''}%
\providecommand \bibnamefont  [1]{#1}%
\providecommand \bibfnamefont [1]{#1}%
\providecommand \citenamefont [1]{#1}%
\providecommand \href@noop [0]{\@secondoftwo}%
\providecommand \href [0]{\begingroup \@sanitize@url \@href}%
\providecommand \@href[1]{\@@startlink{#1}\@@href}%
\providecommand \@@href[1]{\endgroup#1\@@endlink}%
\providecommand \@sanitize@url [0]{\catcode `\\12\catcode `\$12\catcode
  `\&12\catcode `\#12\catcode `\^12\catcode `\_12\catcode `\%12\relax}%
\providecommand \@@startlink[1]{}%
\providecommand \@@endlink[0]{}%
\providecommand \url  [0]{\begingroup\@sanitize@url \@url }%
\providecommand \@url [1]{\endgroup\@href {#1}{\urlprefix }}%
\providecommand \urlprefix  [0]{URL }%
\providecommand \Eprint [0]{\href }%
\providecommand \doibase [0]{http://dx.doi.org/}%
\providecommand \selectlanguage [0]{\@gobble}%
\providecommand \bibinfo  [0]{\@secondoftwo}%
\providecommand \bibfield  [0]{\@secondoftwo}%
\providecommand \translation [1]{[#1]}%
\providecommand \BibitemOpen [0]{}%
\providecommand \bibitemStop [0]{}%
\providecommand \bibitemNoStop [0]{.\EOS\space}%
\providecommand \EOS [0]{\spacefactor3000\relax}%
\providecommand \BibitemShut  [1]{\csname bibitem#1\endcsname}%
\let\auto@bib@innerbib\@empty
\bibitem [{\citenamefont {Reddy}\ \emph {et~al.}(1998)\citenamefont {Reddy},
  \citenamefont {Prakash},\ and\ \citenamefont
  {Lattimer}}]{Reddy.Prakash.Lattimer:1998}%
  \BibitemOpen
  \bibfield  {author} {\bibinfo {author} {\bibfnamefont {S.}~\bibnamefont
  {Reddy}}, \bibinfo {author} {\bibfnamefont {M.}~\bibnamefont {Prakash}}, \
  and\ \bibinfo {author} {\bibfnamefont {J.~M.}\ \bibnamefont {Lattimer}},\
  }\href {\doibase 10.1103/PhysRevD.58.013009} {\bibfield  {journal} {\bibinfo
  {journal} {Phys. Rev. D}\ }\textbf {\bibinfo {volume} {58}},\ \bibinfo
  {pages} {013009} (\bibinfo {year} {1998})}\BibitemShut {NoStop}%
\bibitem [{\citenamefont {Janka}\ \emph {et~al.}(2007)\citenamefont {Janka},
  \citenamefont {Langanke}, \citenamefont {Marek}, \citenamefont
  {Mart{\'i}nez-Pinedo},\ and\ \citenamefont
  {M{\"u}ller}}]{Janka.Langanke.ea:2007}%
  \BibitemOpen
  \bibfield  {author} {\bibinfo {author} {\bibfnamefont {H.-T.}\ \bibnamefont
  {Janka}}, \emph{et al.},\ }\href
  {\doibase 10.1016/j.physrep.2007.02.002} {\bibfield  {journal} {\bibinfo
  {journal} {Phys. Repts.}\ }\textbf {\bibinfo {volume} {442}},\ \bibinfo
  {pages} {38} (\bibinfo {year} {2007})}\BibitemShut {NoStop}%
\bibitem [{\citenamefont {{Hirata}}\ \emph {et~al.}(1987)\citenamefont
  {{Hirata}}, \citenamefont {{Kajita}}, \citenamefont {{Koshiba}},
  \citenamefont {{Nakahata}},\ and\ \citenamefont
  {{Oyama}}}]{Hirata.Kajita.ea:1987}%
  \BibitemOpen
  \bibfield  {author} {\bibinfo {author} {\bibfnamefont {K.}~\bibnamefont
  {{Hirata}}}, \emph{et al.},\ }\href@noop {}
  {\bibfield  {journal} {\bibinfo  {journal} {Phys. Rev. Lett.}\ }\textbf
  {\bibinfo {volume} {58}},\ \bibinfo {pages} {1490} (\bibinfo {year}
  {1987})}\BibitemShut {NoStop}%
\bibitem [{\citenamefont {Koshiba}(1992)}]{Koshiba:1992}%
  \BibitemOpen
  \bibfield  {author} {\bibinfo {author} {\bibfnamefont {M.}~\bibnamefont
  {Koshiba}},\ }\href {\doibase 10.1016/0370-1573(92)90083-C} {\bibfield
  {journal} {\bibinfo  {journal} {Phys. Repts.}\ }\textbf {\bibinfo {volume}
  {220}},\ \bibinfo {pages} {229} (\bibinfo {year} {1992})}\BibitemShut
  {NoStop}%
\bibitem [{\citenamefont {Bethe}\ and\ \citenamefont
  {Wilson}(1985)}]{Bethe.Wilson:1985}%
  \BibitemOpen
  \bibfield  {author} {\bibinfo {author} {\bibfnamefont {H.~A.}\ \bibnamefont
  {Bethe}}\ and\ \bibinfo {author} {\bibfnamefont {J.~R.}\ \bibnamefont
  {Wilson}},\ }\href {\doibase 10.1086/163343} {\bibfield  {journal} {\bibinfo
  {journal} {Astrophys. J.}\ }\textbf {\bibinfo {volume} {295}},\ \bibinfo
  {pages} {14} (\bibinfo {year} {1985})}\BibitemShut {NoStop}%
\bibitem [{\citenamefont {{Mueller}}\ \emph {et~al.}(2012)\citenamefont
  {{Mueller}}, \citenamefont {{Janka}},\ and\ \citenamefont
  {{Marek}}}]{Mueller.Janka.Marek:2012}%
  \BibitemOpen
  \bibfield  {author} {\bibinfo {author} {\bibfnamefont {B.}~\bibnamefont
  {{Mueller}}}, \bibinfo {author} {\bibfnamefont {H.-T.}\ \bibnamefont
  {{Janka}}}, \ and\ \bibinfo {author} {\bibfnamefont {A.}~\bibnamefont
  {{Marek}}},\ }\href@noop {} {\bibfield  {journal} {\bibinfo  {journal} {ArXiv
  e-prints}\ } (\bibinfo {year} {2012})},\ \Eprint
  {http://arxiv.org/abs/1202.0815} {arXiv:1202.0815 [astro-ph.SR]} \BibitemShut
  {NoStop}%
\bibitem [{\citenamefont {{Duncan}}\ \emph {et~al.}(1986)\citenamefont
  {{Duncan}}, \citenamefont {{Shapiro}},\ and\ \citenamefont
  {{Wasserman}}}]{duncan.shapiro.wasserman:1986}%
  \BibitemOpen
  \bibfield  {author} {\bibinfo {author} {\bibfnamefont {R.~C.}\ \bibnamefont
  {{Duncan}}}, \bibinfo {author} {\bibfnamefont {S.~L.}\ \bibnamefont
  {{Shapiro}}}, \ and\ \bibinfo {author} {\bibfnamefont {I.}~\bibnamefont
  {{Wasserman}}},\ }\href {\doibase 10.1086/164587} {\bibfield  {journal}
  {\bibinfo  {journal} {Astrophys. J.}\ }\textbf {\bibinfo {volume} {309}},\
  \bibinfo {pages} {141} (\bibinfo {year} {1986})}\BibitemShut {NoStop}%
\bibitem [{\citenamefont {{Qian}}(2003)}]{Qian:2003}%
  \BibitemOpen
  \bibfield  {author} {\bibinfo {author} {\bibfnamefont {Y.-Z.}\ \bibnamefont
  {{Qian}}},\ }\href {\doibase 10.1016/S0146-6410(02)00178-3} {\bibfield
  {journal} {\bibinfo  {journal} {Prog. Part. Nucl. Phys.}\ }\textbf {\bibinfo
  {volume} {50}},\ \bibinfo {pages} {153} (\bibinfo {year} {2003})}\BibitemShut
  {NoStop}%
\bibitem [{\citenamefont {{Duan}}\ \emph {et~al.}(2010)\citenamefont {{Duan}},
  \citenamefont {{Fuller}},\ and\ \citenamefont
  {{Qian}}}]{Duan.Fuller.Qian:2010}%
  \BibitemOpen
  \bibfield  {author} {\bibinfo {author} {\bibfnamefont {H.}~\bibnamefont
  {{Duan}}}, \bibinfo {author} {\bibfnamefont {G.~M.}\ \bibnamefont
  {{Fuller}}}, \ and\ \bibinfo {author} {\bibfnamefont {Y.}~\bibnamefont
  {{Qian}}},\ }\href {\doibase 10.1146/annurev.nucl.012809.104524} {\bibfield
  {journal} {\bibinfo  {journal} {Ann. Rev. Nucl. Part. Sci.}\ }\textbf
  {\bibinfo {volume} {60}},\ \bibinfo {pages} {569} (\bibinfo {year}
  {2010})}\BibitemShut {NoStop}%
\bibitem [{\citenamefont {{Woosley}}\ \emph {et~al.}(1990)\citenamefont
  {{Woosley}}, \citenamefont {{Hartmann}}, \citenamefont {{Hoffman}},\ and\
  \citenamefont {{Haxton}}}]{Woosley.Hartmann.ea:1990}%
  \BibitemOpen
  \bibfield  {author} {\bibinfo {author} {\bibfnamefont {S.~E.}\ \bibnamefont
  {{Woosley}}}, \emph{et al.},\ }\href {\doibase 10.1086/168839} {\bibfield  {journal}
  {\bibinfo  {journal} {Astrophys. J.}\ }\textbf {\bibinfo {volume} {356}},\
  \bibinfo {pages} {272} (\bibinfo {year} {1990})}\BibitemShut {NoStop}%
\bibitem [{\citenamefont {Heger}\ \emph {et~al.}(2005)\citenamefont {Heger},
  \citenamefont {Kolbe}, \citenamefont {Haxton}, \citenamefont {Langanke},
  \citenamefont {Mart{\'\i}nez-Pinedo},\ and\ \citenamefont
  {Woosley}}]{Heger.Kolbe.ea:2005}%
  \BibitemOpen
  \bibfield  {author} {\bibinfo {author} {\bibfnamefont {A.}~\bibnamefont
  {Heger}}, \emph{et al.},\ }\href@noop {}
  {\bibfield  {journal} {\bibinfo  {journal} {Phys. Lett. B}\ }\textbf
  {\bibinfo {volume} {606}},\ \bibinfo {pages} {258} (\bibinfo {year}
  {2005})}\BibitemShut {NoStop}%
\bibitem [{\citenamefont {Banerjee}\ \emph {et~al.}(2011)\citenamefont
  {Banerjee}, \citenamefont {Haxton},\ and\ \citenamefont
  {Qian}}]{Banerjee.Haxton.Qian:2011}%
  \BibitemOpen
  \bibfield  {author} {\bibinfo {author} {\bibfnamefont {P.}~\bibnamefont
  {Banerjee}}, \bibinfo {author} {\bibfnamefont {W.~C.}\ \bibnamefont
  {Haxton}}, \ and\ \bibinfo {author} {\bibfnamefont {Y.-Z.}\ \bibnamefont
  {Qian}},\ }\href {\doibase 10.1103/PhysRevLett.106.201104} {\bibfield
  {journal} {\bibinfo  {journal} {Phys. Rev. Lett.}\ }\textbf {\bibinfo
  {volume} {106}},\ \bibinfo {pages} {201104} (\bibinfo {year}
  {2011})}\BibitemShut {NoStop}%
\bibitem [{\citenamefont {{Keil}}\ \emph {et~al.}(2003)\citenamefont {{Keil}},
  \citenamefont {{Raffelt}},\ and\ \citenamefont
  {{Janka}}}]{keil.raffelt.janka:2003}%
  \BibitemOpen
  \bibfield  {author} {\bibinfo {author} {\bibfnamefont {M.~T.}\ \bibnamefont
  {{Keil}}}, \bibinfo {author} {\bibfnamefont {G.~G.}\ \bibnamefont
  {{Raffelt}}}, \ and\ \bibinfo {author} {\bibfnamefont {H.-T.}\ \bibnamefont
  {{Janka}}},\ }\href {\doibase 10.1086/375130} {\bibfield  {journal} {\bibinfo
   {journal} {Astrophys. J.}\ }\textbf {\bibinfo {volume} {590}},\ \bibinfo
  {pages} {971} (\bibinfo {year} {2003})}\BibitemShut {NoStop}%
\bibitem [{\citenamefont {Fischer}\ \emph {et~al.}(2012)\citenamefont
  {Fischer}, \citenamefont {Mart{\'i}nez-Pinedo}, \citenamefont {Hempel},\ and\
  \citenamefont {Liebend{\"o}rfer}}]{Fischer.Martinez-Pinedo.ea:2012}%
  \BibitemOpen
  \bibfield  {author} {\bibinfo {author} {\bibfnamefont {T.}~\bibnamefont
  {Fischer}}, \emph{et al.},\ }\href {\doibase 10.1103/PhysRevD.85.083003} {\bibfield
   {journal} {\bibinfo  {journal} {Phys. Rev. D}\ }\textbf {\bibinfo {volume}
  {85}},\ \bibinfo {pages} {083003} (\bibinfo {year} {2012})}\BibitemShut
  {NoStop}%
\bibitem [{\citenamefont {Woosley}\ \emph {et~al.}(1994)\citenamefont
  {Woosley}, \citenamefont {Wilson}, \citenamefont {Mathews}, \citenamefont
  {Hoffman},\ and\ \citenamefont {Meyer}}]{Woosley.Wilson.ea:1994}%
  \BibitemOpen
  \bibfield  {author} {\bibinfo {author} {\bibfnamefont {S.~E.}\ \bibnamefont
  {Woosley}}, \emph{et al.},\ }\href {\doibase
  10.1086/174638} {\bibfield  {journal} {\bibinfo  {journal} {Astrophys. J.}\
  }\textbf {\bibinfo {volume} {433}},\ \bibinfo {pages} {229} (\bibinfo {year}
  {1994})}\BibitemShut {NoStop}%
\bibitem [{\citenamefont {Takahashi}\ \emph {et~al.}(1994)\citenamefont
  {Takahashi}, \citenamefont {Witti},\ and\ \citenamefont
  {Janka}}]{Takahashi.Witti.Janka:1994}%
  \BibitemOpen
  \bibfield  {author} {\bibinfo {author} {\bibfnamefont {K.}~\bibnamefont
  {Takahashi}}, \bibinfo {author} {\bibfnamefont {J.}~\bibnamefont {Witti}}, \
  and\ \bibinfo {author} {\bibfnamefont {H.-T.}\ \bibnamefont {Janka}},\
  }\href@noop {} {\bibfield  {journal} {\bibinfo  {journal} {Astron. \&
  Astrophys.}\ }\textbf {\bibinfo {volume} {286}},\ \bibinfo {pages} {857}
  (\bibinfo {year} {1994})}\BibitemShut {NoStop}%
\bibitem [{\citenamefont {{Hoffman}}\ \emph {et~al.}(1997)\citenamefont
  {{Hoffman}}, \citenamefont {{Woosley}},\ and\ \citenamefont
  {{Qian}}}]{hoffman.woosley.qian:1997}%
  \BibitemOpen
  \bibfield  {author} {\bibinfo {author} {\bibfnamefont {R.~D.}\ \bibnamefont
  {{Hoffman}}}, \bibinfo {author} {\bibfnamefont {S.~E.}\ \bibnamefont
  {{Woosley}}}, \ and\ \bibinfo {author} {\bibfnamefont {Y.-Z.}\ \bibnamefont
  {{Qian}}},\ }\href {\doibase 10.1086/304181} {\bibfield  {journal} {\bibinfo
  {journal} {Astrophys. J.}\ }\textbf {\bibinfo {volume} {482}},\ \bibinfo
  {pages} {951} (\bibinfo {year} {1997})}\BibitemShut {NoStop}%
\bibitem [{\citenamefont {Liebend{\"o}rfer}\ \emph {et~al.}(2001)\citenamefont
  {Liebend{\"o}rfer}, \citenamefont {Mezzacappa}, \citenamefont {Thielemann},
  \citenamefont {{Bronson Messer}}, \citenamefont {{Raphael Hix}},\ and\
  \citenamefont {Bruenn}}]{Liebendoerfer.Mezzacappa.ea:2001a}%
  \BibitemOpen
  \bibfield  {author} {\bibinfo {author} {\bibfnamefont {M.}~\bibnamefont
  {Liebend{\"o}rfer}}, \emph{et al.},\ }\href {\doibase 10.1103/PhysRevD.63.103004}
  {\bibfield  {journal} {\bibinfo  {journal} {Phys. Rev. D}\ }\textbf {\bibinfo
  {volume} {63}},\ \bibinfo {eid} {103004} (\bibinfo {year}
  {2001})}\BibitemShut {NoStop}%
\bibitem [{\citenamefont {Buras}\ \emph {et~al.}(2006)\citenamefont {Buras},
  \citenamefont {Rampp}, \citenamefont {Janka},\ and\ \citenamefont
  {Kifonidis}}]{Buras.Rampp.ea:2006}%
  \BibitemOpen
  \bibfield  {author} {\bibinfo {author} {\bibfnamefont {R.}~\bibnamefont
  {Buras}}, \emph{et al.},\ }\href
  {\doibase 10.1051/0004-6361:20053783} {\bibfield  {journal} {\bibinfo
  {journal} {Astron. \& Astrophys.}\ }\textbf {\bibinfo {volume} {447}},\
  \bibinfo {pages} {1049} (\bibinfo {year} {2006})}\BibitemShut {NoStop}%
\bibitem [{\citenamefont {{H{\"u}depohl}}\ \emph {et~al.}(2010)\citenamefont
  {{H{\"u}depohl}}, \citenamefont {{M{\"u}ller}}, \citenamefont {{Janka}},
  \citenamefont {{Marek}},\ and\ \citenamefont
  {{Raffelt}}}]{Huedepohl.Mueller.ea:2010}%
  \BibitemOpen
  \bibfield  {author} {\bibinfo {author} {\bibfnamefont {L.}~\bibnamefont
  {{H{\"u}depohl}}}, \emph{et al.},\
  }\href {\doibase 10.1103/PhysRevLett.104.251101} {\bibfield  {journal}
  {\bibinfo  {journal} {Phys. Rev. Lett.}\ }\textbf {\bibinfo {volume} {104}},\
  \bibinfo {eid} {251101} (\bibinfo {year} {2010})}\BibitemShut {NoStop}%
\bibitem [{\citenamefont {Fischer}\ \emph {et~al.}(2010)\citenamefont
  {Fischer}, \citenamefont {Whitehouse}, \citenamefont {Mezzacappa},
  \citenamefont {Thielemann},\ and\ \citenamefont
  {Liebend\"orfer}}]{Fischer.Whitehouse.ea:2010}%
  \BibitemOpen
  \bibfield  {author} {\bibinfo {author} {\bibfnamefont {T.}~\bibnamefont
  {Fischer}}, \emph{et al.},\ }\href {\doibase 10.1051/0004-6361/200913106} {\bibfield
  {journal} {\bibinfo  {journal} {Astron. \& Astrophys.}\ }\textbf {\bibinfo
  {volume} {517}},\ \bibinfo {pages} {A80} (\bibinfo {year}
  {2010})}\BibitemShut {NoStop}%
\bibitem [{\citenamefont {Fr{\"o}hlich}\ \emph {et~al.}(2006)\citenamefont
  {Fr{\"o}hlich}, \citenamefont {Mart{\'\i}nez-Pinedo}, \citenamefont
  {Liebend{\"o}rfer}, \citenamefont {Thielemann}, \citenamefont {Bravo},
  \citenamefont {Hix}, \citenamefont {Langanke},\ and\ \citenamefont
  {Zinner}}]{Froehlich.Martinez-Pinedo.ea:2006}%
  \BibitemOpen
  \bibfield  {author} {\bibinfo {author} {\bibfnamefont {C.}~\bibnamefont
  {Fr{\"o}hlich}}, \emph{et al.},\ }\href {\doibase
  10.1103/PhysRevLett.96.142502} {\bibfield  {journal} {\bibinfo  {journal}
  {Phys. Rev. Lett.}\ }\textbf {\bibinfo {volume} {96}},\ \bibinfo {eid}
  {142502} (\bibinfo {year} {2006})}\BibitemShut {NoStop}%
\bibitem [{\citenamefont {Pruet}\ \emph {et~al.}(2006)\citenamefont {Pruet},
  \citenamefont {Hoffman}, \citenamefont {Woosley}, \citenamefont {Janka},\
  and\ \citenamefont {Buras}}]{Pruet.Hoffman.ea:2006}%
  \BibitemOpen
  \bibfield  {author} {\bibinfo {author} {\bibfnamefont {J.}~\bibnamefont
  {Pruet}}, \emph{et al.},\ }\href {\doibase
  10.1086/503891} {\bibfield  {journal} {\bibinfo  {journal} {Astrophys. J.}\
  }\textbf {\bibinfo {volume} {644}},\ \bibinfo {pages} {1028} (\bibinfo {year}
  {2006})}\BibitemShut {NoStop}%
\bibitem [{\citenamefont {Wanajo}(2006)}]{Wanajo:2006}%
  \BibitemOpen
  \bibfield  {author} {\bibinfo {author} {\bibfnamefont {S.}~\bibnamefont
  {Wanajo}},\ }\href {\doibase 10.1086/505483} {\bibfield  {journal} {\bibinfo
  {journal} {Astrophys. J.}\ }\textbf {\bibinfo {volume} {647}},\ \bibinfo
  {pages} {1323} (\bibinfo {year} {2006})}\BibitemShut {NoStop}%
\bibitem [{\citenamefont {Lattimer}\ and\ \citenamefont
  {Swesty}(1991)}]{Lattimer.Swesty:1991}%
  \BibitemOpen
  \bibfield  {author} {\bibinfo {author} {\bibfnamefont {J.~M.}\ \bibnamefont
  {Lattimer}}\ and\ \bibinfo {author} {\bibfnamefont {F.~D.}\ \bibnamefont
  {Swesty}},\ }\href@noop {} {\bibfield  {journal} {\bibinfo  {journal} {Nucl.
  Phys. A}\ }\textbf {\bibinfo {volume} {535}},\ \bibinfo {pages} {331}
  (\bibinfo {year} {1991})}\BibitemShut {NoStop}%
\bibitem [{\citenamefont {{Shen}}\ \emph {et~al.}(1998)\citenamefont {{Shen}},
  \citenamefont {{Toki}}, \citenamefont {{Oyamatsu}},\ and\ \citenamefont
  {{Sumiyoshi}}}]{Shen.Toki.ea:1998a}%
  \BibitemOpen
  \bibfield  {author} {\bibinfo {author} {\bibfnamefont {H.}~\bibnamefont
  {{Shen}}}, \bibinfo {author} {\bibfnamefont {H.}~\bibnamefont {{Toki}}},
  \bibinfo {author} {\bibfnamefont {K.}~\bibnamefont {{Oyamatsu}}}, \ and\
  \bibinfo {author} {\bibfnamefont {K.}~\bibnamefont {{Sumiyoshi}}},\
  }\href@noop {} {\bibfield  {journal} {\bibinfo  {journal} {Nucl. Phys. A}\
  }\textbf {\bibinfo {volume} {637}},\ \bibinfo {pages} {435} (\bibinfo {year}
  {1998})}\BibitemShut {NoStop}%
\bibitem [{\citenamefont {{Reddy}}\ \emph {et~al.}(1999)\citenamefont
  {{Reddy}}, \citenamefont {{Prakash}}, \citenamefont {{Lattimer}},\ and\
  \citenamefont {{Pons}}}]{Reddy.Prakash.ea:1999}%
  \BibitemOpen
  \bibfield  {author} {\bibinfo {author} {\bibfnamefont {S.}~\bibnamefont
  {{Reddy}}}, \bibinfo {author} {\bibfnamefont {M.}~\bibnamefont {{Prakash}}},
  \bibinfo {author} {\bibfnamefont {J.~M.}\ \bibnamefont {{Lattimer}}}, \ and\
  \bibinfo {author} {\bibfnamefont {J.~A.}\ \bibnamefont {{Pons}}},\ }\href
  {\doibase 10.1103/PhysRevC.59.2888} {\bibfield  {journal} {\bibinfo
  {journal} {Phys. Rev. C}\ }\textbf {\bibinfo {volume} {59}},\ \bibinfo
  {pages} {2888} (\bibinfo {year} {1999})}\BibitemShut {NoStop}%
\bibitem [{\citenamefont {{Burrows}}\ and\ \citenamefont
  {{Sawyer}}(1998)}]{Burrows.Sawyer:1998}%
  \BibitemOpen
  \bibfield  {author} {\bibinfo {author} {\bibfnamefont {A.}~\bibnamefont
  {{Burrows}}}\ and\ \bibinfo {author} {\bibfnamefont {R.~F.}\ \bibnamefont
  {{Sawyer}}},\ }\href {\doibase 10.1103/PhysRevC.58.554} {\bibfield  {journal}
  {\bibinfo  {journal} {Phys. Rev. C}\ }\textbf {\bibinfo {volume} {58}},\
  \bibinfo {pages} {554} (\bibinfo {year} {1998})}\BibitemShut {NoStop}%
\bibitem [{\citenamefont {{Burrows}}\ and\ \citenamefont
  {{Sawyer}}(1999)}]{Burrows.Sawyer:1999}%
  \BibitemOpen
  \bibfield  {author} {\bibinfo {author} {\bibfnamefont {A.}~\bibnamefont
  {{Burrows}}}\ and\ \bibinfo {author} {\bibfnamefont {R.~F.}\ \bibnamefont
  {{Sawyer}}},\ }\href {\doibase 10.1103/PhysRevC.59.510} {\bibfield  {journal}
  {\bibinfo  {journal} {Phys. Rev. C}\ }\textbf {\bibinfo {volume} {59}},\
  \bibinfo {pages} {510} (\bibinfo {year} {1999})}\BibitemShut {NoStop}%
\bibitem [{\citenamefont {Roberts}\ \emph {et~al.}(2012)\citenamefont
  {Roberts}, \citenamefont {Shen}, \citenamefont {Cirigliano}, \citenamefont
  {Pons}, \citenamefont {Reddy},\ and\ \citenamefont
  {Woosley}}]{Roberts.Shen.ea:2012}%
  \BibitemOpen
  \bibfield  {author} {\bibinfo {author} {\bibfnamefont {L.~F.}\ \bibnamefont
  {Roberts}}, \emph{et al.},\ }\href {\doibase
  10.1103/PhysRevLett.108.061103} {\bibfield  {journal} {\bibinfo  {journal}
  {Phys. Rev. Lett.}\ }\textbf {\bibinfo {volume} {108}},\ \bibinfo {pages}
  {061103} (\bibinfo {year} {2012})}\BibitemShut {NoStop}%
\bibitem [{\citenamefont {Bruenn}(1985)}]{Bruenn:1985}%
  \BibitemOpen
  \bibfield  {author} {\bibinfo {author} {\bibfnamefont {S.~W.}\ \bibnamefont
  {Bruenn}},\ }\href {\doibase 10.1086/191056} {\bibfield  {journal} {\bibinfo
  {journal} {Astrophys. J. Suppl.}\ }\textbf {\bibinfo {volume} {58}},\
  \bibinfo {pages} {771} (\bibinfo {year} {1985})}\BibitemShut {NoStop}%
\bibitem [{\citenamefont {Langanke}\ and\ \citenamefont
  {Mart{\'\i}nez-Pinedo}(2003)}]{Langanke.Martinez-Pinedo:2003}%
  \BibitemOpen
  \bibfield  {author} {\bibinfo {author} {\bibfnamefont {K.}~\bibnamefont
  {Langanke}}\ and\ \bibinfo {author} {\bibfnamefont {G.}~\bibnamefont
  {Mart{\'\i}nez-Pinedo}},\ }\href@noop {} {\bibfield  {journal} {\bibinfo
  {journal} {Rev. Mod. Phys.}\ }\textbf {\bibinfo {volume} {75}},\ \bibinfo
  {pages} {819} (\bibinfo {year} {2003})}\BibitemShut {NoStop}%
\bibitem [{\citenamefont {{Timmes}}\ and\ \citenamefont
  {{Arnett}}(1999)}]{Timmes.Arnett:1999}%
  \BibitemOpen
  \bibfield  {author} {\bibinfo {author} {\bibfnamefont {F.~X.}\ \bibnamefont
  {{Timmes}}}\ and\ \bibinfo {author} {\bibfnamefont {D.}~\bibnamefont
  {{Arnett}}},\ }\href {\doibase 10.1086/313271} {\bibfield  {journal}
  {\bibinfo  {journal} {Astrophys. J. Suppl.}\ }\textbf {\bibinfo {volume}
  {125}},\ \bibinfo {pages} {277} (\bibinfo {year} {1999})}\BibitemShut
  {NoStop}%
\bibitem [{\citenamefont {Woosley}\ \emph {et~al.}(2002)\citenamefont
  {Woosley}, \citenamefont {Heger},\ and\ \citenamefont
  {Weaver}}]{Woosley.Heger.Weaver:2002}%
  \BibitemOpen
  \bibfield  {author} {\bibinfo {author} {\bibfnamefont {S.~E.}\ \bibnamefont
  {Woosley}}, \bibinfo {author} {\bibfnamefont {A.}~\bibnamefont {Heger}}, \
  and\ \bibinfo {author} {\bibfnamefont {T.~A.}\ \bibnamefont {Weaver}},\
  }\href@noop {} {\bibfield  {journal} {\bibinfo  {journal} {Rev. Mod. Phys.}\
  }\textbf {\bibinfo {volume} {74}},\ \bibinfo {pages} {1015} (\bibinfo {year}
  {2002})}\BibitemShut {NoStop}%
\bibitem [{\citenamefont {Dasgupta}\ \emph {et~al.}(2009)\citenamefont
  {Dasgupta}, \citenamefont {Dighe}, \citenamefont {Raffelt},\ and\
  \citenamefont {Smirnov}}]{Dasgupta.Dighe.ea:2009}%
  \BibitemOpen
  \bibfield  {author} {\bibinfo {author} {\bibfnamefont {B.}~\bibnamefont
  {Dasgupta}}, \emph{et al.},\ }\href
  {\doibase 10.1103/PhysRevLett.103.051105} {\bibfield  {journal} {\bibinfo
  {journal} {Phys. Rev. Lett.}\ }\textbf {\bibinfo {volume} {103}},\ \bibinfo
  {pages} {051105} (\bibinfo {year} {2009})}\BibitemShut {NoStop}%
\bibitem [{\citenamefont {{Hoffman}}\ \emph {et~al.}(1996)\citenamefont
  {{Hoffman}}, \citenamefont {{Woosley}}, \citenamefont {{Fuller}},\ and\
  \citenamefont {{Meyer}}}]{Hoffman.Woosley.ea:1996}%
  \BibitemOpen
  \bibfield  {author} {\bibinfo {author} {\bibfnamefont {R.~D.}\ \bibnamefont
  {{Hoffman}}}, \emph{et al.},\ }\href {\doibase 10.1086/176986} {\bibfield  {journal} {\bibinfo
   {journal} {Astrophys. J.}\ }\textbf {\bibinfo {volume} {460}},\ \bibinfo
  {pages} {478} (\bibinfo {year} {1996})}\BibitemShut {NoStop}%
\bibitem [{\citenamefont {{Qian}}\ and\ \citenamefont
  {{Woosley}}(1996)}]{Qian.Woosley:1996}%
  \BibitemOpen
  \bibfield  {author} {\bibinfo {author} {\bibfnamefont {Y.-Z.}\ \bibnamefont
  {{Qian}}}\ and\ \bibinfo {author} {\bibfnamefont {S.~E.}\ \bibnamefont
  {{Woosley}}},\ }\href {\doibase 10.1086/177973} {\bibfield  {journal}
  {\bibinfo  {journal} {Astrophys. J.}\ }\textbf {\bibinfo {volume} {471}},\
  \bibinfo {pages} {331} (\bibinfo {year} {1996})}\BibitemShut {NoStop}%
\bibitem [{\citenamefont {{Lattimer}}\ and\ \citenamefont
  {{Lim}}(2012)}]{Lattimer.Lim:2012}%
  \BibitemOpen
  \bibfield  {author} {\bibinfo {author} {\bibfnamefont {J.~M.}\ \bibnamefont
  {{Lattimer}}}\ and\ \bibinfo {author} {\bibfnamefont {Y.}~\bibnamefont
  {{Lim}}},\ }\href@noop {} {\  (\bibinfo {year} {2012})},\ \Eprint
  {http://arxiv.org/abs/1203.4286} {arXiv:1203.4286 [nucl-th]} \BibitemShut
  {NoStop}%
\end{thebibliography}

%

\end{document}